\documentclass[aip,jcp,reprint]{revtex4-1}
\usepackage{epsfig}
\begin{document}

\title{Local elastic response measured near the colloidal glass transition}

\author{D. Anderson$^{1}$, D. Schaar$^{1}$, 
H.~G.~E.~Hentschel$^{1}$,
J. Hay$^{1}$,
Piotr Habdas$^{2}$, and Eric R. Weeks$^{1}$}
\affiliation{$^{1}$ Department of Physics, Emory University,
Atlanta, GA 30322\\$^{2}$ Department of Physics, Saint Joseph's University, Philadelphia, PA 19131}
\date{\today}

\begin{abstract}
  We examine the response of a dense colloidal suspension to
  a local force applied by a small magnetic bead.  For small
  forces, we find a linear relationship between the force
  and the displacement, suggesting the medium is elastic, even
  though our colloidal samples macroscopically behave as fluids.
  We interpret this as a measure of the strength of colloidal
  caging, reflecting the proximity of the samples' volume
  fractions to the colloidal glass transition.  The strain field
  of the colloidal particles surrounding the magnetic probe
  appears similar to that of an isotropic homogeneous elastic
  medium.  When the applied force is removed,
  the strain relaxes as a stretched exponential in
  time.  We introduce a model that suggests this behavior is
  due to the diffusive relaxation of strain in the colloidal sample.
\end{abstract}

\maketitle

\section{Introduction}

Glass is an amorphous solid:  despite the lack of long-range
order, a glassy material is elastic rather than viscous.  As a
glass-forming material is cooled, its viscosity rises dramatically
by many orders of magnitude; flow becomes difficult and slow.
The elastic behavior of a glass, then, could be reframed as the
material being probed on time scales too quickly for liquid-like
flow to occur.  The origins of this elasticity and the nature of
the glass transition are a quite active area of research
\cite{dyre06,schuh07,lubchenko07,cavagna09,berthier11rmp}.

A related question is to what extent the macroscopic elastic
behavior extends to the microscopic scale.  For example,
simulations have seen that the elastic moduli are spatially
heterogeneous in some cases \cite{leonforte06,tsamados09}.
Certainly the macroscopic elastic response treats the material as
a continuum, whereas on a scale of the constituent molecules this
could be a poor approximation.

Colloids are a simple model system which can be used to study
the glass transition and the properties of glassy materials
\cite{sciortino05,pusey08,hunter12rpp}.  Colloidal suspensions are
composed of solid particles in a liquid.  The particles diffuse
due to Brownian motion, but this diffusion is impeded at high
particle concentration.  In many experiments, colloidal particles
only have short-range repulsive interactions, and the particles
can be approximated as hard spheres \cite{pusey86,royall12}.
In such samples the control parameter is the volume fraction
$\phi$ and glasses are found for $\phi > \phi_g \approx 0.58$
\cite{pusey86,hunter12rpp}.  Near the glass transition, the
viscosity rises quite dramatically, although some evidence
suggests that perhaps it truly diverges at a point above $\phi_g$
\cite{cheng02}.  In particular, an alternative divergence point
is ``random close packing'' (rcp).  This is the largest volume
fraction possible for a sample that is still amorphously packed
(rather than crystalline) \cite{torquato00}.  The volume fraction
$\phi_{rcp}$ is known through simulations and taken to be
$\approx 0.64$, which agrees with early experiments done with
ball bearings \cite{bernal64}.

Of course, the presence of the continuous liquid surrounding
the colloidal particles is important for understanding the flow
properties of colloids.  Approaching the colloidal glass transition
from $\phi < \phi_g$, samples are viscoelastic \cite{mason95glass}.
Their properties are described by both viscous and elastic moduli
which are frequency-dependent.  The viscosity mentioned above is
understood to be the low-frequency limit.  Indeed, if a glass is
defined as an amorphous solid -- a sample that does not flow --
then it is important to recognize that whether or not it flows
depends on the time scale of observation \cite{hunter12rpp}.

While macroscopically one considers viscous and elastic behavior,
microscopically one considers diffusion.
A molecule or tracer particle in a
fluid sample has a nonzero diffusion constant, which decreases
to zero as the glass transition is approached.  The decrease of
the diffusion coefficient is attributed to caging.  On short time
scales, particles diffuse in a ``cage'' formed by their neighbors.
On longer time scales, these cages rearrange and particles
can move throughout the sample.  As the glass transition is
approached, the cage rearrangements occur less frequently, thus
decreasing the diffusivity \cite{rabani97,doliwa98,weeks02sub}.
These cages provide a sort of elasticity for individual particles
\cite{schweizer03,schweizer07,sussman11}.  Particles which
try to move away from the centers of their cage experience
a restoring force \cite{doliwa98,weeks02sub,schweizer03}.
If a constant external force is exerted on a particle, it
will slowly move through the colloidal suspension as cages
rearrange, although the force needs to be kept small when
$\phi$ is close to $\phi_g$ to avoid nonlinear behavior
\cite{williams06,gazuz09,gnann11,winter12,gnann12}.
If a sufficiently high external force is applied to
a particle, it can break the cages and move through
the sample more freely (typically with a nonlinear
relationship between the force and resulting velocity)
\cite{hastings03,habdas04,carpen05,reichhardt06,olsonreichhardt10,gnann11,winter12},
disturbing and rearranging particles as it moves
\cite{hastings03,drocco05}.

In this paper we probe the elastic response of a dense colloidal
suspension by locally exerting a small force on a magnetic bead.
Our studies are conducted on a time scale faster than the sample
can relax due to diffusion.  We find that the sample responds
elastically, with a Young's modulus $E$ that rises as the glass
transition is approached, and a Poisson ratio $\sigma$ equal to 1/2.
We also study the relaxation of the magnetic bead after it has
been displaced and the force removed.  This relaxation is faster
for higher volume fraction samples.  We present a model that
captures the stretched exponential character of the relaxation,
by assuming that in our colloidal sample
stress diffuses away to infinity at long times.

\section{Experimental Methods}
\label{methods}


The colloidal suspensions are made of poly-(methylmethacrylate)
particles, sterically stabilized by a thin layer of
poly-12-hydroxystearic acid \cite{antl86}.  The particles have
a radius $a=1.55$~$\mu$m, a polydispersity of $\sim$5\%, and
are dyed with rhodamine \cite{dinsmore01}.  
The uncertainty of the mean particle
radius is $\pm 0.01$~$\mu$m.  The particles are slightly
charged, but their glass transition is still at $\phi_g = 0.58
\pm 0.01$, signaled by the diffusion of particles going to zero on
experimental time scales (several hours).  The colloidal particles
are suspended in a mixture of cyclohexylbromide/{\it cis}- and
{\it trans}- decalin which nearly matches both the density and the
index of refraction of the colloidal particles \cite{dinsmore01}.
The density of this solvent is 1.232~g/cm$^3$, the index of
refraction is 1.495, and the viscosity is
$\eta=2.18$~mPa$\cdot$s.
The particles are fluorescently labeled for visualization
\cite{dinsmore01}.
Before beginning experiments, we stir the sample with an air bubble
to break up any pre-existing crystalline regions, then wait 20
minutes before taking data.  


A small quantity of superparamagnetic beads (Dynal M450, coated with
glycidyl ether reactive groups) with a radius of 2.25 $\mu$m are
added to the colloidal suspension.  We do not observe attraction
or repulsion between the colloidal particles and the magnetic
beads, in either dilute or concentrated samples.  The beads are
not completely monodisperse in their magnetic properties, and
our calibration finds the variability in the effective magnetic
force applied to different beads to be less than 10$\%$.  Also,
the magnetic beads are not density matched, and their effective
weight is 0.1 pN.  This is a factor of ten smaller than the smallest
horizontally applied magnetic forces in our experiments.  We study
isolated magnetic beads at least 35 $\mu$m from the sample chamber
boundary and from other magnetic beads.


For some of the data (Figs.~\ref{fig-force},\ref{fig-spring} in
particular), we use a conventional Leica DMIRB inverted microscope
with a CCD camera.  A magnetic bead appears as a large dark circle
in our images and its position as a function of time is found
using standard particle tracking techniques \cite{crocker96}.
For these experiments, the typical image size is $15 \times
40$~$\mu$m$^2$, with images taken at a rate of 30 frames per second.
The magnetic bead position is resolved to within 0.04~$\mu$m;
see Fig.~\ref{fig-force} for example.

For other data, we use a ThermoNoran Oz confocal microscope.
With this microscope, we acquire images of area 80~$\mu$m $\times$
75~$\mu$m at a rate of 30 frames per second ($256 \times 240$
pixels$^2$).  The magnetic beads are not fluorescent and thus
appear black on the background of dyed colloidal particles
(see Fig.~\ref{fig-rawpiv}).  Again, we use particle tracking
procedures to follow the motion of the magnetic particle with a
resolution of 0.04~$\mu$m.  With either the confocal microscope or
the video microscope, we only track the motion of particles in 2D:
for the experiments in this paper, the magnetic particle position
always remains with the imaging plane of the microscope.

We additionally use the confocal microscope to take
three-dimensional images of these samples to determine the volume
fraction \cite{dinsmore01}.  We count the particles within the
imaged volume, and convert from the measured number density $n$
to volume fraction $\phi$ using $\phi = (4\pi/3)na^3$.  Due to our
uncertainty of the mean value of the particle radius $a$, we have a
systematic uncertainty of 2\% of our $\phi$ values \cite{poon12}.
That is, our $\phi$ values are fairly accurate when compared with
each other, but a reported value of $\phi=0.50$ is uncertain by
$\pm 0.01$.


We use a strong Neodymium permanent magnet mounted on a micrometer
positioner to control the force applied to the magnetic beads.
The micrometer accurately reproduces the magnet position and
thus our uncertainty in the force between different experiments
is limited by the magnetic bead variability, rather than the
magnet positioning.  For many experiments, a computer-controlled
stepper motor attached to the micrometer allows us to slowly and
controllably vary the applied force over two orders of magnitude.

For other experiments, we want to apply a short-duration magnetic
force.  We mount the magnet on a linear actuator (Ultra Motion
D-A.25-HT17).  The magnet is then brought close to the magnetic
bead resulting in a high force acting on the magnetic bead for a
short time; the details are discussed below.

Our experiments are controlled-force experiments, in contrast to
controlled displacement \cite{squires05,khair06}.  For example,
this means that particles do not necessarily need to move in the
direction of the applied force, although we discuss below that
they appear to always do so.  Prior work by other groups used laser
tweezers to move probe particles through colloidal suspensions at
a controllable velocity \cite{meyer06,wilson09,sriram10,wilson11},
finding many interesting results such as anisotropy of the disturbed
region around the moving particle \cite{meyer06} and a decoupling
of structural and hydrodynamic influences on the particle motion
\cite{wilson11}.  These prior experiments studied probe particles
as they moved over long distances, in contrast to our experiments
described below where the magnetic bead always remains close to
its equilibrium position.

\section{Linear Elastic Response}

In our earlier work, we applied a constant force and observed the
steady-state motion of the magnetic bead \cite{habdas04}.  For
large forces, the velocity of the magnetic bead grew nonlinearly
with increasing force, consistent with shear-thinning.  We also
observed that the velocity was essentially zero below a threshold
force.  The threshold force grew dramatically as the glass
transition was approached \cite{habdas04}.

\begin{figure}
\centerline{\epsfxsize=8.0truecm \epsffile{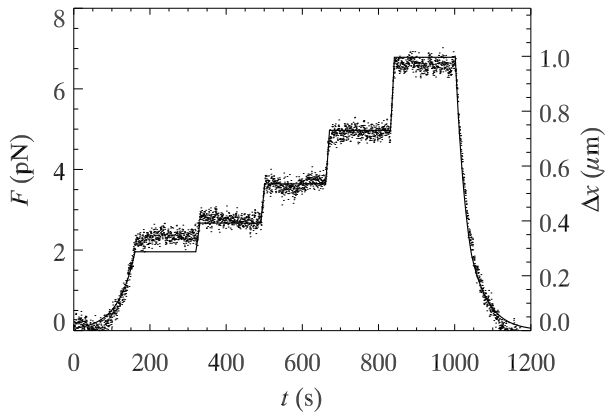}}
\smallskip
\caption{The solid line indicates the applied force as a function
of time.  The points show the measured
displacement $\Delta x$ of the magnetic
bead.  The volume fraction is $\phi=0.55$.
}
\label{fig-force}
\end{figure}

To test the behavior of the sample below the threshold force for
motion, we vary the magnetic force within a range of forces
that are below the threshold force for motion.  To ensure that
the forces used are below the threshold force for motion, each
increase in the force was followed by a waiting period of 160 s to
observe the subsequent motion of the magnetic bead.  The applied
force as a function of time is shown in Fig.~\ref{fig-force} by
the solid line.  In this situation, the magnetic bead has a finite
displacement, rather than a finite velocity; the displacement is
shown by the dots in Fig.~\ref{fig-force}.  As the figure shows,
by appropriately rescaling the units, the displacement of the
magnetic bead from its original position is linearly proportional
to the applied force.  Additionally, during each of the pauses
at constant force, the magnetic bead exhibits slight fluctuations
around its equilibrium position.  Some of this is due to Brownian
motion, and some of this is due to the uncertainty in identifying
the position of the magnetic bead ($\sim 0.04$ $\mu$m).

%
%
\smallskip

\begin{figure}
\centerline{\epsfxsize=8.0truecm \epsffile{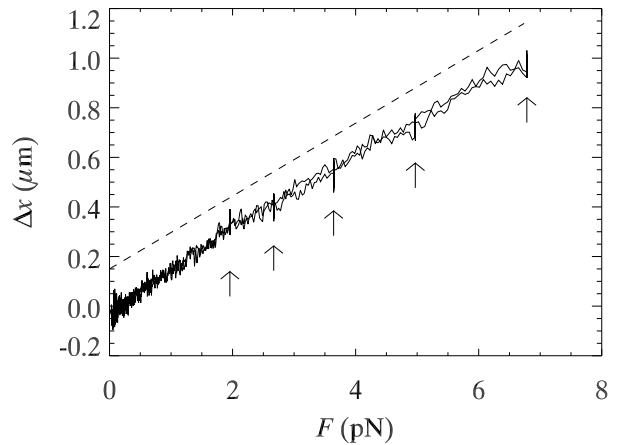}}
\smallskip
\caption{The measured displacement $\Delta x$ of the magnetic
bead as the force is varied; the data correspond to
Fig.~\ref{fig-force}.
The dashed line is a fit to the data,
with the slope leading to an effective spring
constant $k = 6.8 \pm 0.1$ pN/$\mu$m,
offset vertically for clarity.  The arrows indicate locations
where the force was held constant for 160 s (see
Fig.~\ref{fig-force}).
}
\label{fig-spring}
\end{figure}

Further evidence for the linearity is seen in
Fig.~\ref{fig-spring}, showing the displacement data plotted
as a function of the applied force.  The data are linearly
related, and a fit to the data leads to an effective spring
constant $k = 6.8$ pN/$\mu$m.   The arrows shown in the figure
indicate locations where the magnetic force was constant.
This spring constant is quite large, equivalent to 4000 $k_B
T / a^2$ using the radius of the colloidal particles or 8400
$k_B T / a_{MB}^2$ using the radius of the magnetic bead.

%
%
\smallskip

\section{Particle Displacement Fields}
\label{experiment}

Unfortunately, the results of Figs.~\ref{fig-force},
\ref{fig-spring} are atypical in one important respect:  for many
experiments lasting longer than $O$(100 s), the magnetic bead 
experiences a cage rearrangement and does not return to its original
position.  
The spring constant before and after any such displacement
is always the same, to within our uncertainty.  Accordingly,
to complement the slow experiments of Figs.~\ref{fig-force} and
\ref{fig-spring}, we conduct experiments with intermittent and
short pulses of force to see the instantaneous response of the
sample on a time scale quicker than cage rearrangements.
A nondimensional way to consider this is the modified
Peclet number \cite{habdas04}.  $Pe^*$ is the ratio of the time
it would take for a colloidal particle to diffuse its own size to
the time scale for the perturbation.  The diffusive time scale
is $\sim$400-5000~s for the samples we study ($\phi > 0.4$),
and the perturbation time scale is 0.25~s.  Thus, $Pe^* \approx
2000-20000$, signifying that Brownian motion is unimportant
on the time scales we study:  particles do not substantially
rearrange their positions during our experiment.  The recovery to
the perturbation (discussed in Sec.~\ref{decay}) takes $O$(10~s),
which is still in the high $Pe$ limit.

%
%
\smallskip
\begin{figure}
\centerline{\epsfxsize=8.0truecm \epsffile{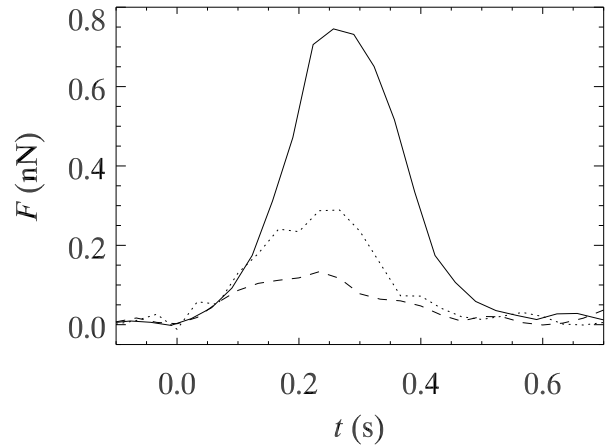}}
\smallskip
\caption{Applied force as a function of time, for the three largest
maximum forces (see Table I).  For all curves, $F(t) = 0$ for $t<0$~s.}
\label{fig-cal}
\end{figure}

To discover the origin of the linear restoring force, we examine
the response of the colloidal
particles surrounding the magnetic bead.  To produce
reproducible initial strains, we attach the external
permanent magnet to a linear actuator as described in
Sec.~\ref{methods}, and move the magnet
toward the sample and then away at maximum speed.  The
force applied is ramped up to a maximum value and then just
as rapidly reduced.  To calibrate the force as a function
of time, $F(t)$, we use this procedure to exert a force
on magnetic beads suspended in glycerol.  Such beads move
with velocity $v(t)$, from which we deduce the
force $F(t)$ using Stokes' Law, $F = 6 \pi \eta a_{\rm MB}
v$, with the viscosity of glycerol $\eta = 0.934$~Pa$\cdot$s.
The resulting $F(t)$ data are plotted in Fig.~\ref{fig-cal}
for the three largest $F_{\rm max}$.  These correspond to the
cases where the external magnet is moved the closest to the
sample.  For example, the top curve in Fig.~\ref{fig-cal}
takes longer to reach its peak and longer to return
to $F = 0$, as the magnet has farther to move.

%
%
\begin{table}
\caption{
The five different maximum forces applied, and the integrated
impulse $I = \int F(t) dt$.  The calibration procedure (described
in the text) has
an intrinsic $F_{\rm max}$ uncertainty of $\pm 0.05$ nN and an
$I$ uncertainty of $\pm 0.005$ nN$\cdot$s.
Due to variability between different
magnetic beads, for a given magnetic bead there is also an
overall systematic uncertainty of $\pm$10\%.
Graphs of $F(t)$ for the three
largest values of $F_{\rm max}$ are shown in
Fig.~\protect\ref{fig-cal}.
}
\begin{tabular}{l|l}
$F_{\rm max}$ (nN) & $I$ (nN s)\\
\hline
0.042 & 0.011\\
0.077  & 0.020\\
0.13  & 0.036\\
0.29   & 0.068\\
0.75   & 0.17\\
\end{tabular}
\label{forcetable}
\end{table}

The peaks of the curves, $F_{\rm max}$, are confirmed by
measuring the velocity of magnetic beads in glycerol while the
external magnet is fixed in its closest distance to the sample
for a given forcing protocol.  Those results agree quite well
with the values measured from the $F(t)$ data.  In the results
that follow, we refer to the different $F(t)$ by their maximum
values $F_{\max}$ which are listed in Table~\ref{forcetable}.
An additional way to quantify the $F(t)$ is by their time
integral, $I=\int F(t) dt$, yielding an impulse which is
applied to the magnetic bead.  These values are also listed
in Table~\ref{forcetable}.  In each case, the ratio $I / F_{\rm
max} = 0.25 \pm 0.03$ s, suggesting that our choice of
using $F_{\rm max}$ is correctly representing $I$ as well, and
that the effective pulse duration is a quarter of a second.  This
time scale is short compared to the Brownian time scale $a_{\rm
MB}^2/D_{\rm MB} = 110$~s.

%
%
\smallskip
\begin{figure}
\centerline{\epsfxsize=3.5truecm \epsffile{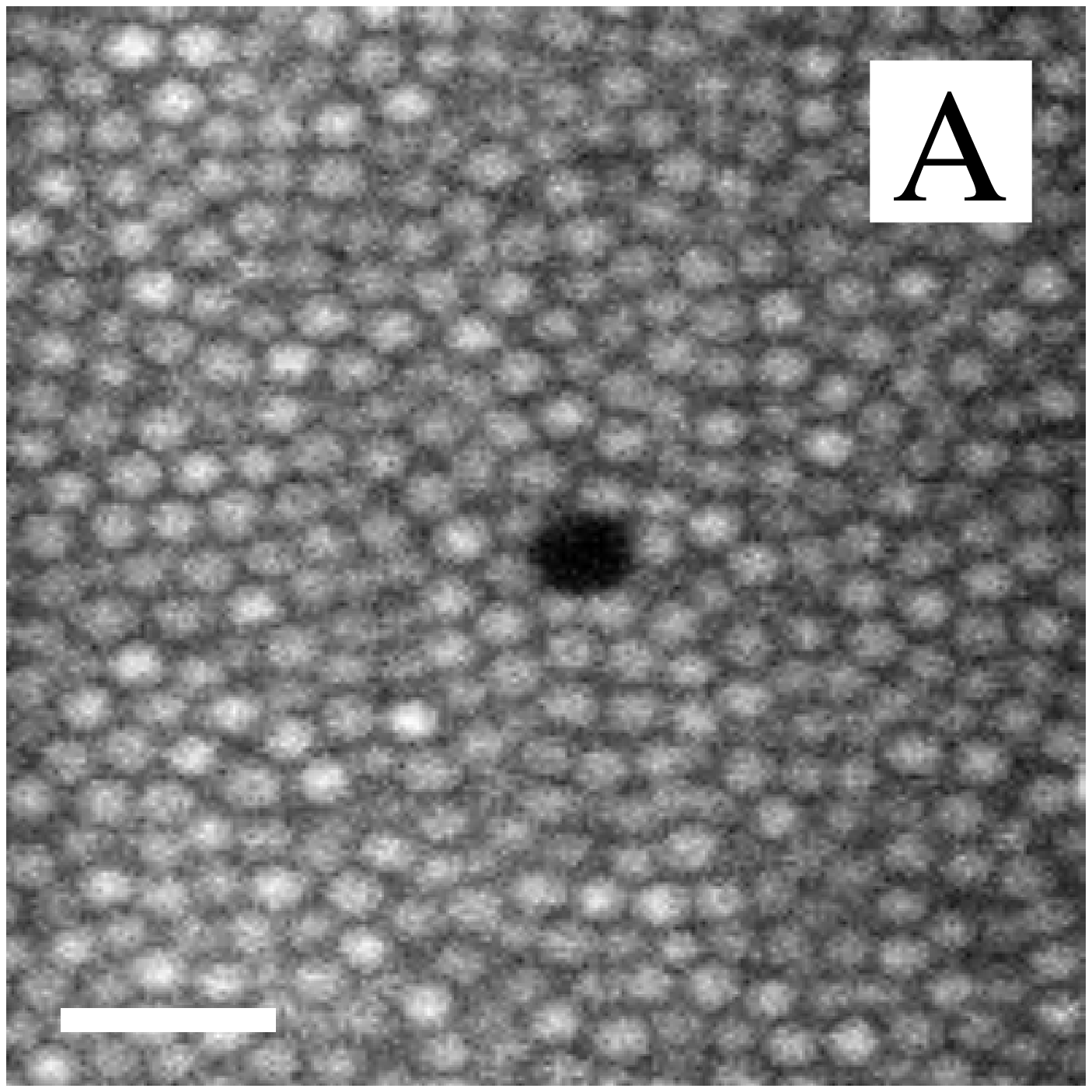}
\epsfxsize=3.5truecm \epsffile{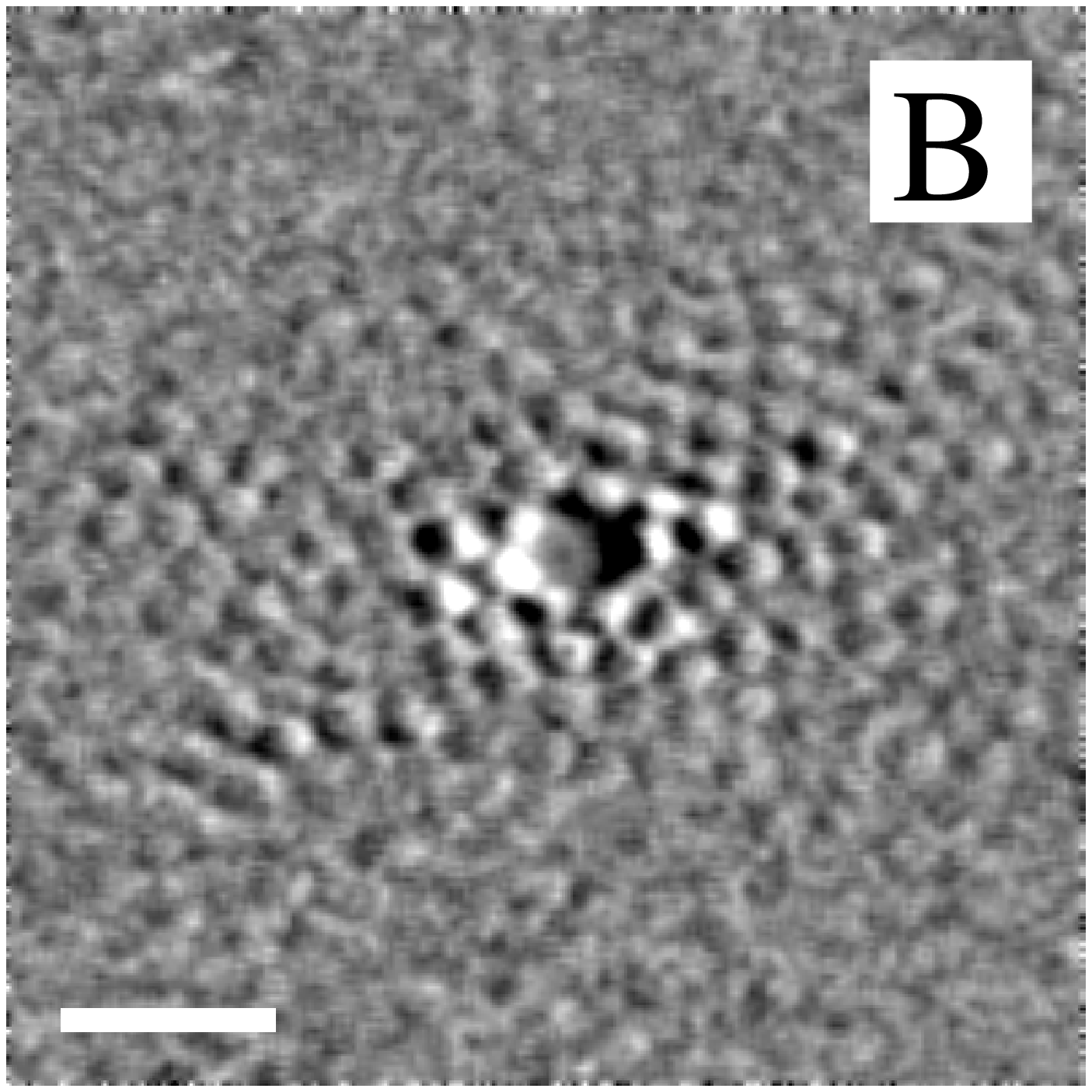}
} \centerline{
\epsfxsize=3.5truecm \epsffile{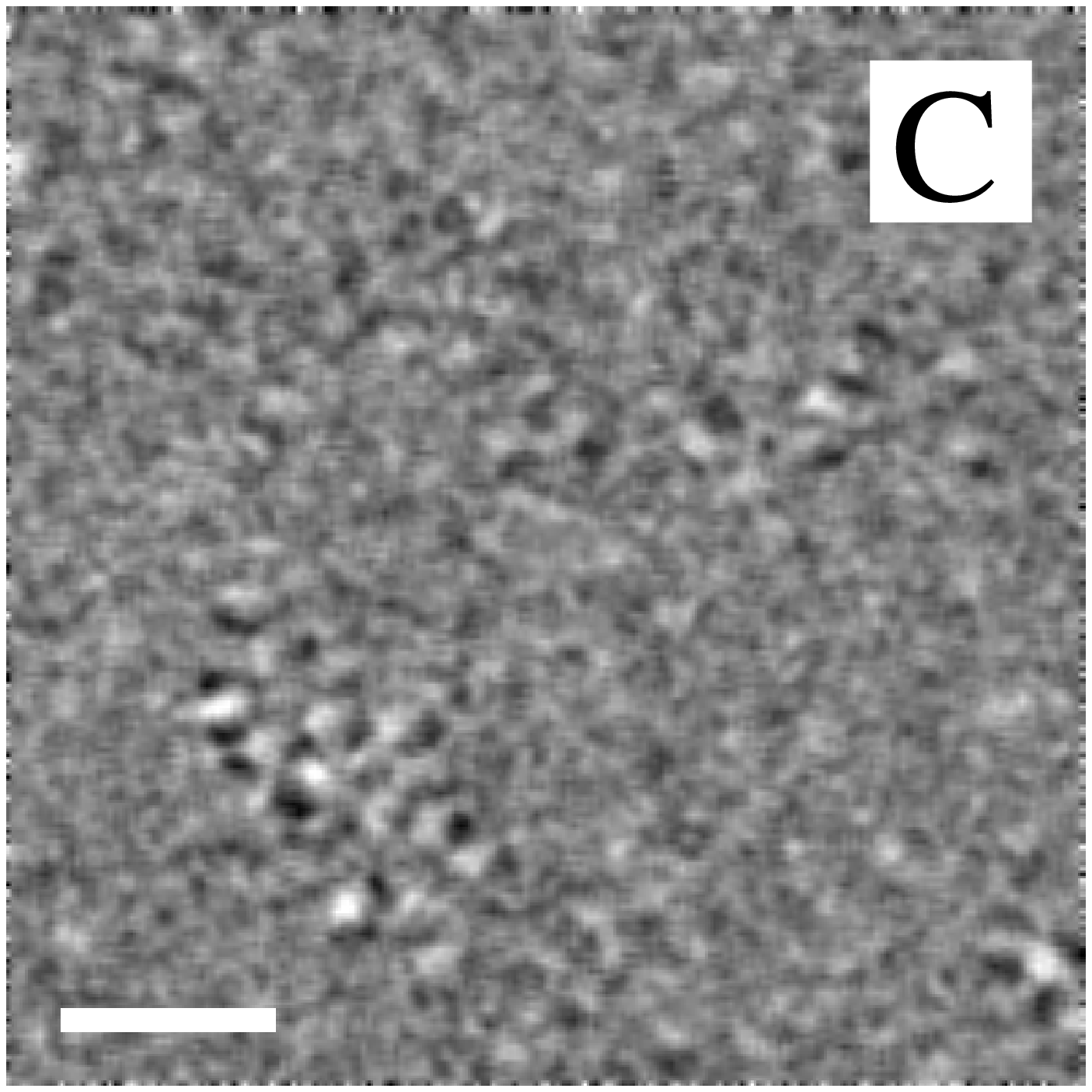}
\epsfxsize=3.5truecm \epsffile{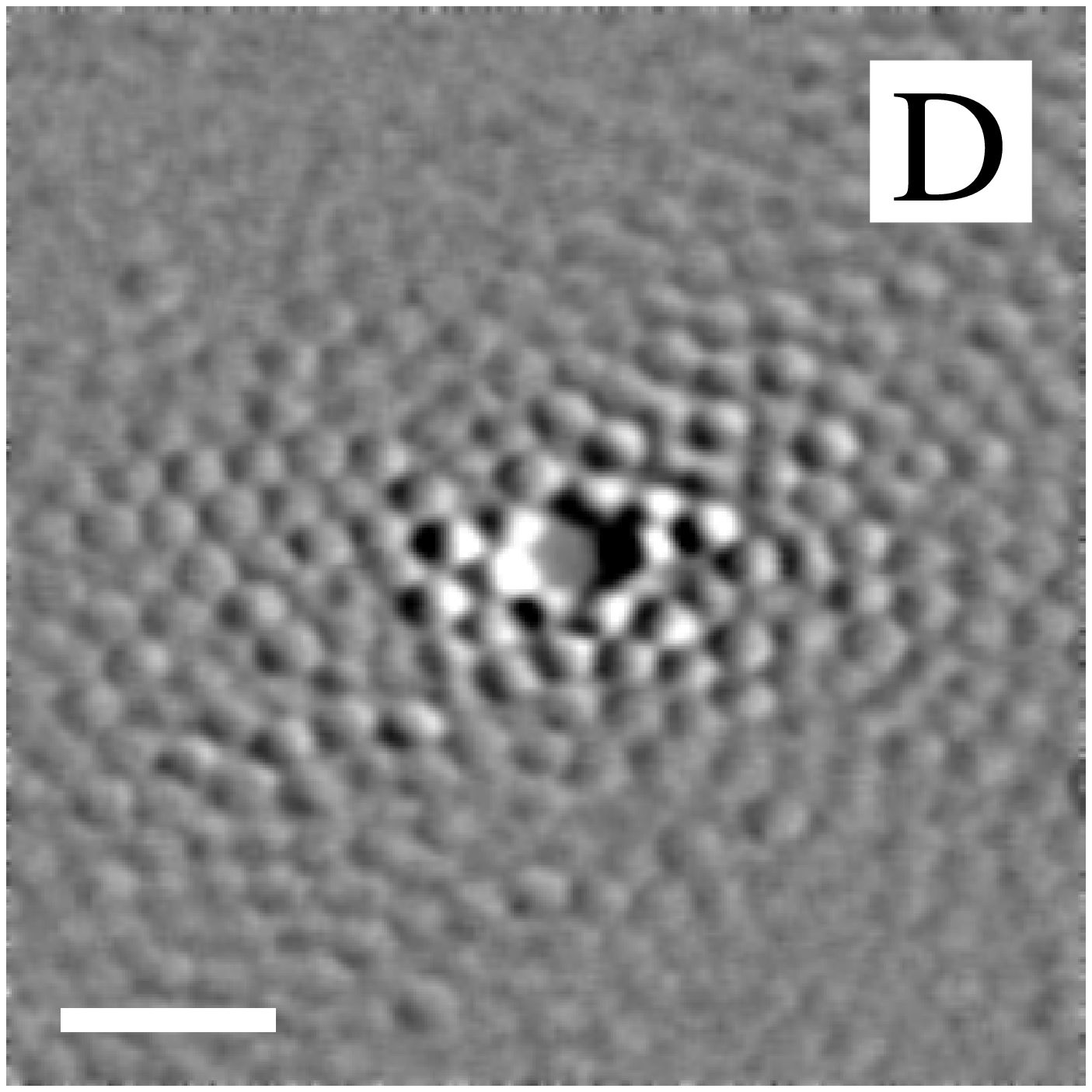}
}
\smallskip
\caption{(a) Raw image of particles, before the force is
applied.  
(b) Difference between ``before'' and ``after'' a single force pulse is
applied.  
(c) Difference between two ``after'' images
for two subsequent pulses.
(d)  As
the images are reproducible, a sequence of eight ``before''
pictures are averaged together, and likewise eight ``after''
pictures.  This picture is the difference between these average
images.
For all pictures, the scale bar is 10 microns long, 
the volume fraction is $\phi = 0.49$,
and the applied
force is $F_{\rm max} = 0.29$ nN.
}
\label{fig-rawpiv}
\end{figure}

The response of the sample to a pulse is shown in
Fig.~\ref{fig-rawpiv}(b).  This is a difference image
formed by subtracting the raw image before the pulse [such
as Fig.~\ref{fig-rawpiv}(a)] from the raw image after the
pulse.  In this case, the magnetic particle has moved to the
left, as indicated by the white crescent on its left side.
Because the magnetic bead is black and the colloids are white,
the colloidal motion is indicated by the direction of the
black crescents, and is also clearly leftward.  Furthermore,
the overall disturbed region of colloids is a fairly smooth
function of space.  Adjacent colloids move similar distances
in Fig.~\ref{fig-rawpiv}(b).

This displacement field is highly reproducible, as is shown by
creating a difference image in Fig.~\ref{fig-rawpiv}(c) between
two images both taken when the colloids are maximally displaced
from their equilibrium positions.  The difference image is nearly
completely gray, showing that the displacements are virtually
the same.  Slight local variations are due to Brownian motion of
particles within their cages, but no rearrangements occur over
the duration of the experiment.  Due to the reproducibility of
the experiment, it is reasonable to average the images ``before''
and the images ``after'' to reduce the variability caused by
Brownian motion.  The resulting difference image is shown in
Fig.~\ref{fig-rawpiv}(d) and emphasizes the smoothly spatially
varying displacements of the colloids.  It can be seen that the
applied force is not exactly on the $x$-axis of the image; this
is dealt with in the analysis below.

%
%
\smallskip
\begin{figure}
\centerline{\epsfxsize=7truecm \epsffile{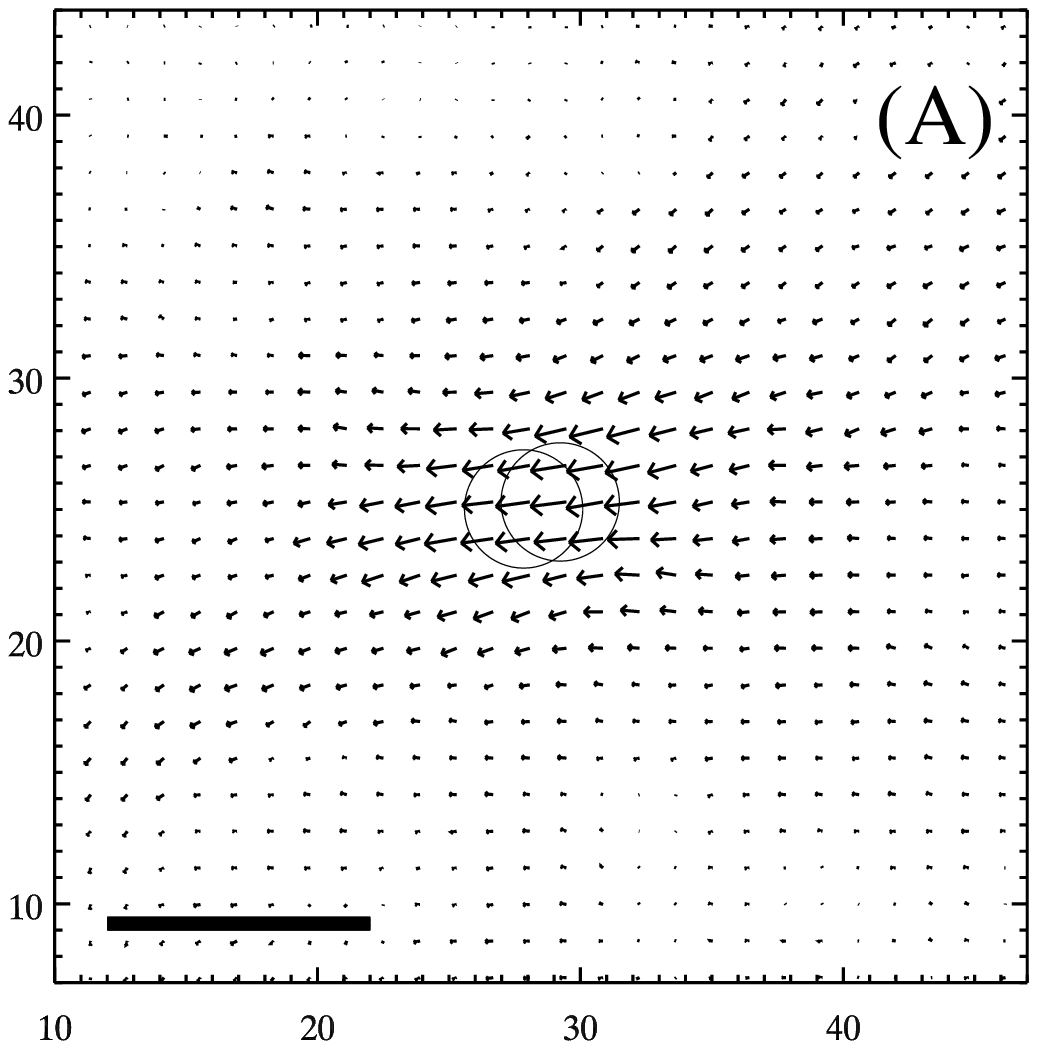}}
\centerline{\epsfxsize=7truecm \epsffile{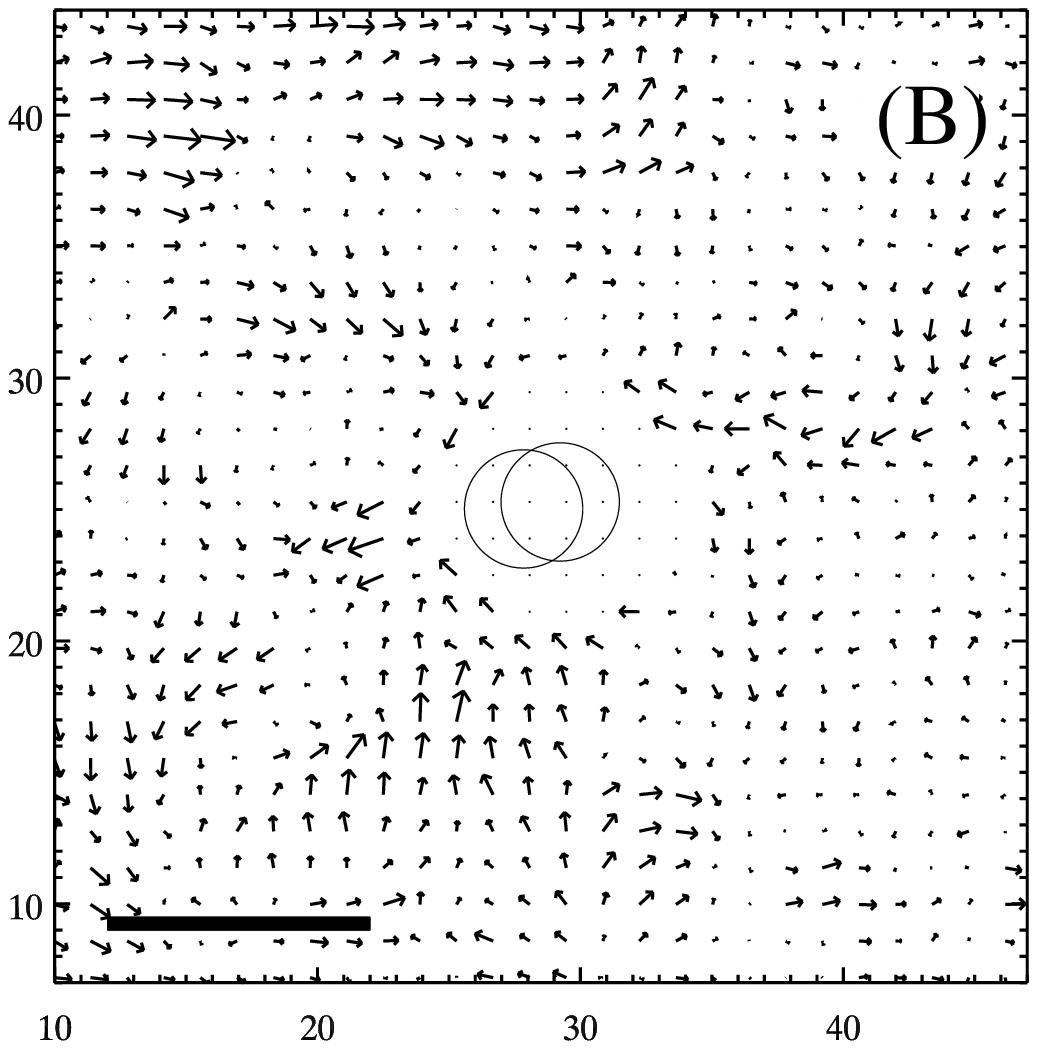}}
\smallskip
\caption{(a) Displacement field based on data shown in
Fig.~\protect\ref{fig-rawpiv}(b).  The arrows indicate displacements
of the colloidal particles.  (b) Residual displacement field after
subtracting off the fit to Eqn.~\ref{formula}.  The arrows are
magnified by a factor of 5; in reality, the longest displacement
vectors in panel (b) are 0.3~$\mu$m.  The central region near the
magnetic bead is removed for clarity.  For both panels, the circles
indicate the initial and final positions of the magnetic bead, which
moved from right to left, and are drawn to scale.  The scale bar
is 10 microns long.  The data correspond to Figs.~\ref{fig-piv}
and \ref{fig-fit}: $\phi=0.49$ and $F_{\rm max} = 0.29$ nN.
Note that a displacement vector is calculated for every pixel in
the raw images; here only every 6th vector is drawn.
}
\label{fig-piv}
\end{figure}

To quantify the images shown in Fig.~\ref{fig-rawpiv}, we
perform particle image velocimetry (PIV) on the pairs of
``before'' and ``after'' images.  This method is frequently
used in experimental fluid mechanics, and does not depend on
identifying or tracking individual particles \cite{weeks10}.  A small
window in the first image is taken, and cross-correlated with
the same size window in the second image.  By moving the second
window around in the second image, we find which piece
of the second image is best correlated with the piece from the
first image.  The shift required for this maximum correlation
is a displacement vector reflecting how the particles have
moved between the two images, and in particular represents
the displacement vector for the center of the window.  We use
a window size that roughly encompasses two particles, although
our results are not sensitive to this choice.  The technique
is merely correlating the images and the particles provide
contrast to help this work.  
A typical displacement field is
shown in Fig.~\ref{fig-piv}(a), corresponding to the images shown
in Fig.~\ref{fig-rawpiv}(a,b).  
For the PIV analysis, we use the individual raw
images such as Fig.~\ref{fig-rawpiv}(a) which leads to
Fig.~\ref{fig-rawpiv}(b), rather than the averaged images which
lead to Fig.~\ref{fig-rawpiv}(d).  After computing the PIV
analysis for each individual pulse, we average the PIV fields
over the pulse sequence for a given $F_{\rm max}$ and $\phi$ to
do the subsequent analysis.

The smoothly varying appearance of the displacement field seen in
Fig.~\ref{fig-piv}(a) suggests trying to fit the strain field to a
simple functional form.  As noted previously, the response of the
magnetic bead to a constant force is a simple linear function,
and so a natural choice is to treat the colloidal suspension as
a homogeneous elastic medium.  In such a medium, the strain field
$\vec{u}$ around a point force $\vec{F}$ applied at the origin is
given by \cite{landau86}:
\begin{equation}
\vec{u} =
{ {1} \over {8 \pi E} }
{ {1+\sigma} \over {1 - \sigma} }
\left[
{ {(3-4\sigma)\vec{F} + \hat{n}(\hat{n} \cdot \vec{F} )} \over {r} }
\right]
\label{originallandau}
\end{equation}
where $E$ is the Young's modulus, $\sigma$ is the Poisson ratio, and
$\hat{n}$ is a unit vector pointing away from the origin.  Using
$\hat{n} = \hat{x}\cos \theta + \hat{y}\sin\theta$ and
$\vec{F}=F\hat{x}$, the equation can be rewritten as:
\begin{equation}
\vec{u} =
{ {1} \over {16 \pi r} }
\left( { {F} \over {E} } \right)
\left( { {1+\sigma} \over {1 - \sigma} } \right)
\left[
{(7-8\sigma + \cos 2\theta)\hat{x} + (\sin 2\theta)\hat{y}}
\right]
\label{formula}
\end{equation}
which highlights the key spatial dependence of the strain field:
it decays as $1/r$, and has a periodic dependence on $2\theta$,
due to the symmetry of the problem about the $x$-axis.
$\theta=0$ corresponds to the direction of the force.

%
\smallskip
\begin{figure}
\centerline{\epsfxsize=8truecm \epsffile{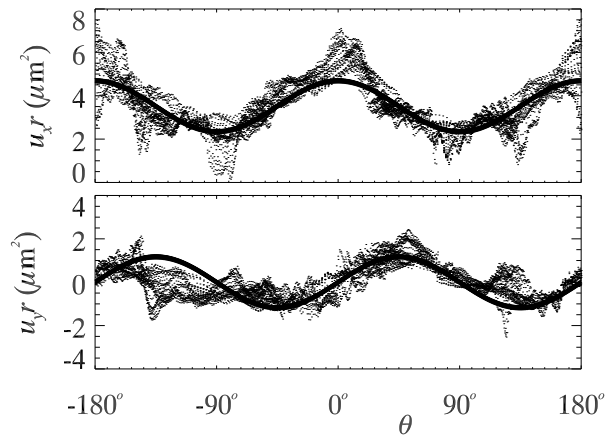}
}
\smallskip
\caption{Rescaled displacement vectors as a function of $\theta$;
compare with Eqn.~\protect\ref{formula}.  The points are the data
and the solid line is the fit to the equation.  The data
correspond to Figs.~\protect\ref{fig-rawpiv} and
\protect\ref{fig-piv}(a), using only data with $r > r_0 = a_{MB}$.
For this fit, $\sigma$ was constrained to be 1/2.
}
\label{fig-fit}
\end{figure}

To test this, we rescale the displacements $u_x$ and $u_y$
(measured from PIV) by $r$.  This collapses the data reasonably
well, as shown in Fig.~\ref{fig-fit}.  Here the data are plotted
as a function of $\theta$, showing the characteristic modulation
in Eqn.~\ref{formula}.  The solid line in Fig.~\ref{fig-fit}
is a fit to the equation.  The amplitude of both fit curves is
constrained by the model to be the same, which is in slight
disagreement with the raw data, where $u_x r$ is somewhat larger
in amplitude than $u_y r$.  This is seen in most of our data sets.
The curve for $u_x r$ is vertically offset, and it can be seen
from Eqn.~\ref{formula} that the magnitude of the offset is
related to the Poisson ratio $\sigma$.

The fit has several parameters.  First, the direction for $\theta=0$
is chosen to be the average direction of all of the displacement
vectors, to correct for the imperfect magnet alignment.  Second,
the two physical parameters to the fit are $\sigma$ and $E$.
A difficulty in determining $E$ is that the true value of $F$ is
unknown:  the model assumes a steady $F$ whereas we apply a pulse.
If the force was held at $F_{\rm max}$, the magnetic bead would
move with a velocity and would not return to its original position,
given the large values of $F_{\rm max}$ we use \cite{habdas04}.
Fortunately, $\vec{u}$ scales reasonably well with $F_{\rm max}$
and thus leads to a consistent value for the Young's modulus $E$,
even if its true magnitude cannot be deduced from our fits.  Third,
we allow for the location of the origin ($x=0,y=0$) to vary.  It is
not obvious if the origin should be at the starting position of the
magnetic bead, the ending position, or elsewhere, especially given
that the magnetic bead is a finite-sized disturbance and the model
assumes a point-sized disturbance.  We adjust the origin so that the
model has the best fit to the data; this typically puts the origin
within 0.3~$\mu$m of the starting position of the magnetic bead.
A final parameter to our fitting algorithm is above what radius
$r_{\rm 0}$ from the magnetic bead the fitting is conducted.
Sufficiently close to the magnetic bead, its finite size begins to
distort the strain field from the model.  We fix $r_0 = a_{MB}$,
and in practice our results are not sensitive to our choice.

%
\smallskip
\begin{figure}
\centerline{\epsfxsize=8truecm \epsffile{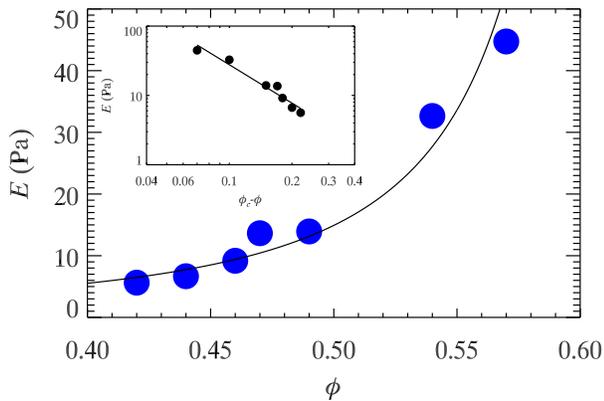}
}
\smallskip
\caption{Young's modulus $E$ as a function of volume fraction
$\phi$.  Due to an inadequately defined applied force, $E$ is
overestimated although this affects all points equally (by a
multiplicative factor) and does not change the shape of the
curve; see the text for a discussion.  The inset shows the
same data plotted as a function of $\phi_c-\phi$ with
$\phi_c=0.64$.  The lines in the main plot and the inset are the
fit to the data using $E = E_0 (\phi_c - \phi)^{-\beta}$ with $E_0 =
0.4$~Pa and $\beta=1.84\pm 0.40$.  The symbol size indicates the
uncertainty.
}
\label{fig-phi}
\end{figure}

Despite the slight disagreements between the raw data and the
model shown in Fig.~\ref{fig-fit}, overall the model is remarkably
successful.  For a few samples, we find that the Young's modulus $E$
increases slightly with increasing $F_{\rm max}$, but more often we
find $E$ is independent of $F_{\rm max}$ and accordingly for each
sample we average $E$ over the different trials with different
$F_{\rm max}$.  The resulting data are
plotted in Fig.~\ref{fig-phi}, and $E$ increases by a factor of
$\sim 8$ as the glass transition is approached.  
Simulations and theory show that elastic
moduli diverge near the jamming transition as $B \sim (\phi_c -
\phi)^{-\beta}$ with $\phi_c \approx 0.64$, the volume fraction of
random close packing \cite{durian95,durian97,ohern03,brito07}.  The exponent
$\beta$ depends on the details of the interparticle interaction
and which modulus is considered.
The inset of Fig.~\ref{fig-phi} shows $E$ plotted as a function
of $(\phi_c-\phi)$ with behavior consistent with a power-law,
although our data extend over only half a decade of $(\phi_c-\phi)$.
Our exponent is $\beta = 1.84 \pm 0.40$, similar to results for
the bulk modulus of hard spheres ($\beta=2$) and shear modulus of
hard spheres ($\beta = 3/2$) \cite{brito07}.

Our values of $E$ are quite similar to those found in a classic
study of viscoelastic shear moduli of colloidal super-cooled liquids
\cite{mason95glass}, although that is a coincidence.  On the one
hand, Eqn.~\ref{formula} assumes the sample is in equilibrium for
the applied force, which is certainly not the case.  Using $F_{\rm
max}$ overestimates $E$.  On the other hand, our particles
are 7.4 times larger than those of Ref.~\cite{mason95glass},
so our moduli should be smaller by a factor of $7.4^3 = 400$.
We can estimate the correct order of magnitude for our data from
Fig.~\ref{fig-spring}, using Eqn.~\ref{formula} with our effective
spring constant $k=6.8$~pN/$\mu$m (setting this equal to $u/F$)
and $r=a_{\rm MB}=2.25$~$\mu$m.  This gives us $E=0.72$~Pa for
$\phi=0.55$, suggesting that our data in Fig.~\ref{fig-phi}
are overestimated by a factor of $O(100)$.  Thus we are in
plausible agreement with the data of Ref.~\cite{mason95glass},
in the high-frequency limit in particular which is most relevant
for our quickly perturbed samples.

An alternative comparison for $E$ can be made with the theory
of Schweizer and Saltzman, who developed an effective ``free
energy'' for a hard sphere trapped in a cage \cite{schweizer03}.
They construct the free energy $F(r)$ as a function of the distance
$r$ from the cage center.  For particle motion within the cage, they
find an effective spring constant depending on $\phi$ as $k \sim k_0
\exp(25.3 \phi)$ with $k_0 = 2.5 \cdot 10^{-4} k_B T/a^2$ (where
$k_B$ is Boltzmann's constant, $T$ is the absolute temperature,
and $a$ is the colloidal particle radius).  In our experiment,
the magnetic bead is larger than the surrounding particles by a
factor of 1.45, so the effective spring constant experienced by
our bead will be larger by $1.45^2=2.1$.  Using $\phi=0.55$
and correcting for the bead size, their 
theory predicts $k \approx 580 kT/a^2$, as compared to our result
of $k=6.8$~pN/$\mu$m$=4000 k_BT/a^2$.  Our result is a factor of
7 larger.  Overall, given the approximations made by the theory
and the uncertainties of the experiment, agreement within a factor
of 7 is suggestive that the origin of the elasticity we observe
is indeed the caging of the particles.

The other key fit parameter in Eqn.~\ref{formula} is the Poisson
ratio $\sigma$.  Over all values of $F_{\rm max}$ and $\phi$
we find $\sigma=0.50 \pm 0.08$.  Values of $\sigma$ larger than
1/2 are unphysical, so we conclude that our data show $\sigma
= 1/2$.  Accordingly, we fix this value and redo the fits to
Eqn.~\ref{formula}, and the values of $E$ that result are the ones
shown in Fig.~\ref{fig-phi} and correspond to the fit curves shown
in Fig.~\ref{fig-fit}.  The physical meaning of $\sigma=1/2$ is
that volume is conserved during deformation: were this sample to
be strained in one direction, the sides would contract sufficient
to conserve volume.  This is plausible, as the sample itself is
an incompressible fluid with solid particles, and additionally
one assumes the volume fraction stays homogeneous during simple
deformations.

As noted above, we allow the direction of the force to be a free
parameter when performing the fit.  This angle is fairly constant,
with a standard deviation of only $4^\circ$ between the different
experiments.  This variability likely reflects measurement error.


The fit shown in Fig.~\ref{fig-fit} is not perfect,
and some systematic deviations from the fit can be seen.
The difference between the fit and the measurements is shown
in Fig.~\ref{fig-piv}(b).  The displacement vectors are
stretched by a factor of 5, and thus greatly exaggerate the
difference.  Nonetheless, this picture looks similar to the
locally nonaffine elastic behaviors seen in some simulations
\cite{tanguy02,wittmer02,leonforte05,leonforte06}
and also images of ``floppy-modes,'' localized normal
modes, and ``soft spots'' known to be present near jamming
\cite{ellenbroek06,brito07jsm,widmercooper08,candelier10,manning11}.
We stress that the majority of the total displacement field shown
in Fig.~\ref{fig-piv}(a) is well-fit by Eqn.~\ref{formula}.

\section{Decay of Strain}
\label{decay}

\subsection{Experimental observations}

\begin{figure}
\centerline{\epsfxsize=8truecm \epsffile{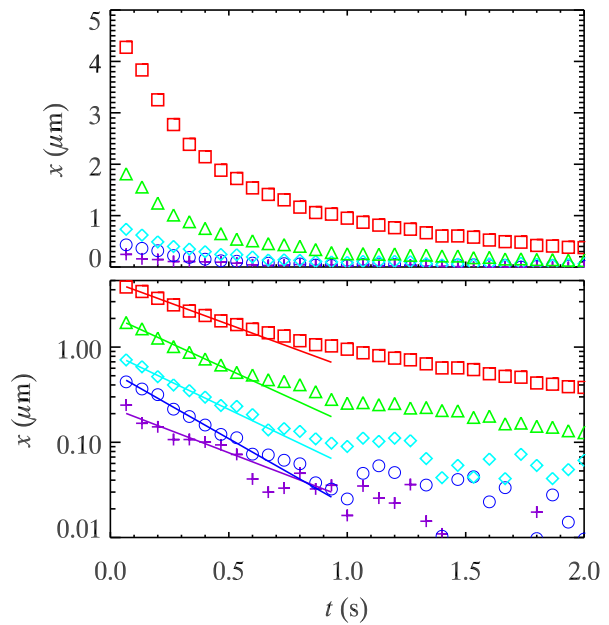}
}
\smallskip
\caption{
Plots of the displacement of the magnetic bead as a function of
time, after the force is removed.  (a) Shows a linear-linear plot
and (b) shows a log-linear plot.
The values of $F_{\rm max}$
are given in Table 1, with the largest initial displacement (red
squares) corresponding to the largest force and the smallest
initial displacement (purple pluses) corresponding to the
smallest force.  In (b), lines are fit to the initial data
($t<0.5$~s) indicating decay time constants of 0.47~s, 0.38~s,
0.37~s, 0.31~s, and 0.46~s (from largest $F_{\rm max}$ to
smallest).
}
\label{fig-relax}
\end{figure}

After the force is removed, the magnetic bead moves
back to its equilibrium position.  Typical data of the
magnetic bead displacement as a function of time are shown in
Fig.~\ref{fig-relax}(a).  Within our resolution, the magnetic bead
is always in the initial position less than 10~s after it starts
the return motion.

%
\smallskip

Figure~\ref{fig-relax}(b) shows the data on a semilog plot,
where straight lines would indicate exponential decay.  While the
initial portion of the data can be fit to straight lines, clear
deviations are seen at longer times.  The decay times found are
$0.3 - 0.5$~s but do not depend systematically on the initial
displacement.  Furthermore, some evidence of memory is seen.
For example, the $F_{\rm max}=0.29$~nN data (green triangles)
go from $x=2.0$ to 0.4~$\mu$m during the time interval $t=0.0$
to 0.7~s.  In contrast, the $F_{\rm max}=0.75$~nN data (red squares)
go from $x=2.0$ to 0.4~$\mu$m during the time interval $t=0.4$ to
1.9~s, taking nearly twice as long to cover the same displacement.
The noisy data seen in Fig.~\ref{fig-relax}(b) at small values
of $x$ are partly due to the uncertainty in determining $x$ ($\pm
0.04$~$\mu$m).  The $x=0$ position is defined by an average at
long times and so is more accurately defined.  Within
our resolution the positions shown in Fig.~\ref{fig-relax}(b)
have not quite decayed to $x=0$ over the time period shown.


%
\smallskip
\begin{figure}
\centerline{\epsfxsize=8truecm \epsffile{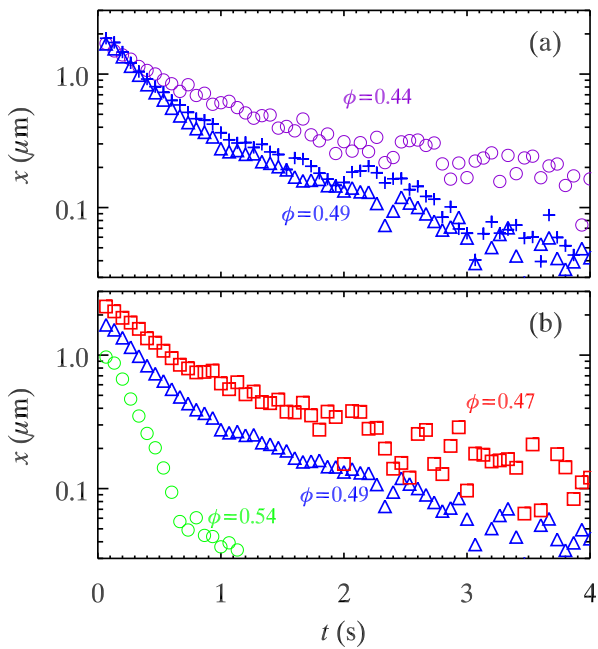}
}
\smallskip
\caption{Relaxation curves for several experiments, demonstrating
that the decay is faster for samples with higher $\phi$.  (a)
Comparison of two samples with $\phi$ as indicated, that have
nearly the same initial displacement.  For the $\phi=0.44$ data, the
force is $F_{\rm max}=0.13$~nN, and for the $\phi=0.49$ data, the
force is $F_{\rm max}=0.29$~nN.  Two different instances are
shown for the $\phi=0.49$ data (triangles and pluses).  
(b) Comparison of three samples with the same force ($F_{\rm
max}=0.29$~nN) but different $\phi$ as indicated.
}
\label{fig-phidecay}
\end{figure}

One trend is that samples with larger $\phi$ (closer
to the glass transition) decay faster, as is shown in
Fig.~\ref{fig-phidecay}.  Given the nontrivial memory effects,
it is not obvious whether to compare data at constant
initial displacement or at constant $F_{\rm max}$.  This distinction
turns out to be unimportant.  Figure \ref{fig-phidecay}(a)
compares two different volume fractions with the same initial
displacement, and the data at larger $\phi$ decay faster.  Figure
\ref{fig-phidecay}(b) shows three different volume fractions with
the same $F_{\rm max}$, with the same trend, data for larger
$\phi$ decay faster.  This makes intuitive sense, as the elastic
modulus is larger for larger $\phi$ (Fig.~\ref{fig-phi}).  While
viscous dissipation rises as the glass transition is approached
\cite{mason95glass}, apparently the elastic contribution to the
magnetic bead relaxation rises faster, resulting in a faster
relaxation.

%
\smallskip
\begin{figure}
\centerline{\epsfxsize=8truecm \epsffile{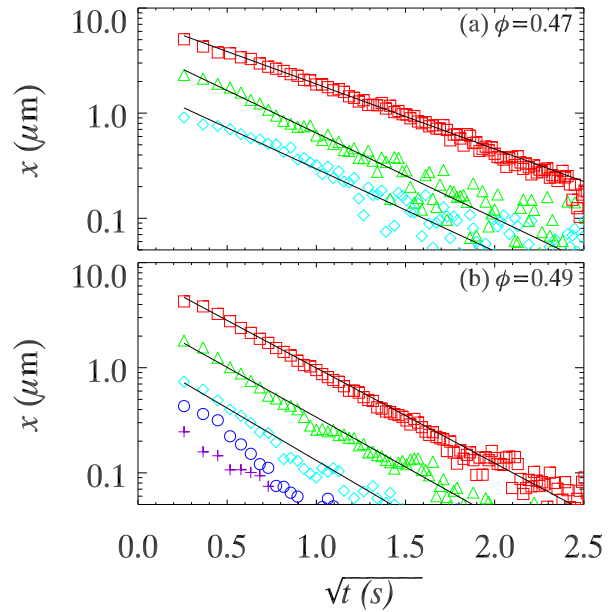}
}
\smallskip
\caption{Displacement plotted as a function of $\sqrt{t}$ for (a)
$\phi=0.47$ and (b) $\phi=0.49$.  The different symbols indicate
different values of $F_{\rm max}$.
The values of $F_{\rm max}$
are given in Table 1, with the largest initial displacement (red
squares) corresponding to the largest force and the smallest
initial displacement (purple pluses) corresponding to the
smallest force.  The straight lines indicate fits to $\sim
\exp(-\sqrt{t/t_0})$.  For (a), the values of $t_0$ are 0.50,
0.29, 0.31~s (top to bottom).  For (b), the values of $t_0$ are
0.23, 0.21, 0.19~s (top to middle).
}
\label{fig-sqrt}
\end{figure}

While the position as a function of time does not appear to
decay exponentially (Fig.~\ref{fig-relax}), plotting the data
in Fig.~\ref{fig-sqrt} as a function of $\sqrt{t}$ suggests
$x \sim \exp(-\sqrt{t/t_0})$.  The value of $t_0$ is slightly
larger for larger initial displacements, although the data in
Fig.~\ref{fig-relax} are fairly parallel within each panel,
showing that $t_0$ is not changing that dramatically.  $t_0$
is clearly larger for lower volume fractions $\phi$.

\subsection{Model of relaxing bead}

To explain the stretched exponential decay process, we develop a
model that treats the relaxation of stresses in the viscoelastic
colloidal sample.  Consider the relaxation dynamics of a magnetic
bead of radius $a_{MB}$ initially at $x=0$ in a viscoelastic
medium that is suddenly displaced at time $t=0$ by an amount $x_0$
due to an imposed force $F$. The bead was originally (at $t<0$)
in equilibrium and the displacement will result both in a force
exerted on the external medium by the bead creating a stress field
in the colloidal medium and a reaction force by this medium on
the bead.  The experiments show clearly the presence of both memory
and a stretched exponential behavior for the bead relaxation. We
show here that if the induced stress field relaxes in a diffusive
manner then such behavior arises. The reaction force will thus in
general be a function of both the applied force $F$ that creates
the inhomogeneous stress field around the bead, as well as time
$t$ due to diffusive relaxation of the stress field. This reaction
force will tend to bring the bead back to its original equilibrium
position due to the elastic forces exerted on the bead together
with a viscoelastic drag force that will dissipate energy. Thus
we can write the equation of motion
\begin{equation}
\label{a}
 m \ddot x = - m \int_0^t\zeta (t-s) \dot x(s) ds - k(F, t) x,
\end{equation}
where the first term on the RHS is the viscoelastic drag on
the bead and the second term represents the elastic force on
the bead in the presence of a relaxing force constant $k(F,
t)$. Because the motion is slow we can replace the viscoelastic
drag by its viscous zero frequency limit. Namely defining $\gamma =
\int_0^\infty \zeta (t) dt$ we can rewrite Eqn.~\ref{a} as
\begin{equation}
\label{b}
 \ddot x = - \gamma \dot x   -\omega (F, t)^2 x.
\end{equation}
We estimate $\gamma = 6 \pi a_{MB} \eta /m$ using Stokes' law
with $\eta$ being the effective viscosity of the colloidal medium,
and we define $\omega (F, t)^2 = k(F ,t)/m$.

Our first challenge is to estimate $k(F, t)$. We can see from
the strain field $u({\bf r})$ induced in the colloidal medium due
to the applied force $F$ (see Eqn.~\ref{originallandau}) that there exists a
length scale $\xi_0$ over which the colloidal displacements will
be greater than the typical colloidal particle radius $a$. This
scale can be estimated as 
\begin{equation}
\label{xi0}
\xi_0 (F) \approx C F/(E a)
\end{equation}
 where $E$
is the Young's modulus of the sample and the constant $C\approx
(1/16 \pi) (1+\sigma)(7-8\sigma)/(1-\sigma ) \approx 0.2$.  Using
our largest $F_{\rm max}=0.75$~nN and $E \approx 15$~Pa, $\xi_0
\approx 4a$.  Given that the medium is not perfectly elastic
but rather viscoelastic, we argue that this region grows
diffusively as the strain is dissipated in the surrounding
medium, leadaing to a growing length scale 
$\xi (F, t)$ where
\begin{equation}
 \label{c}
 \xi (F, t)^2 = \xi_0 (F)^2 +6 D t.
\end{equation}
In Eqn.~\ref{c} $D$ is the diffusion constant characterizing
motion that can relax the strain, which does not require cage
rearrangements.  We discuss $D$ in more detail below.
Using $\xi (F, t)$ we can estimate the
typical strains in the colloidal medium induced by the bead. These
strains $\epsilon \approx x/\xi (F, t)$ reduce with time both due to the
diffusive relaxation of the initial strain and the reduced
imposed forces as the bead returns to its original equilibrium
position. The associated elastic stresses in the colloid are then
$\approx E x/\xi(F, t)$. Thus we are now in a position to
estimate the restoring force on the bead as $k(F, t) x \approx 4
\pi R^2 E x/\xi (F, t)$ or
\begin{equation}
\label{d}
\omega (F, t)^2 \approx 4 \pi R^2 E/(m \xi (F, t)) = \omega_0^2/\sqrt{(1 + t/\tau ) },
\end{equation}
where $\omega_0^2 = 4 \pi R^2 E/(m \xi_0 (F))$ and $\tau = \xi_0(F)^2/6 D$. Substituting Eqn.~\ref{d} into Eqn.~\ref{b} then yields
\begin{equation}
\label{e}
 \ddot x = - \gamma \dot x   -\omega_0^2 x/\sqrt{(1 + t/\tau ) }.
\end{equation}

The relaxational behavior of the bead can now be found by solving
Eqn.~\ref{e} subject to the initial conditions $x (t=0) = x_0$ and
$\dot x(t=0) = 0$.  Though Eqn.~\ref{e} cannot be solved exactly,
it has two limiting forms. For $t \ll \tau$ Eqn.~\ref{e} reduces
to $\ddot x = - \gamma \dot x   -\omega_0^2 x $. This is
the equation of motion for an linear oscillator with two
overdamped modes
\begin{eqnarray}
\label{f}
x(t)  =  (x_0/2) \{&& \alpha_{+} \exp{[-\gamma \alpha_{-} t/2]}\nonumber \\
+&& \alpha_{-} \exp{[-\gamma \alpha_{+} t/2]}\},
\end{eqnarray}
using $\alpha_{\pm} = 1 \pm \sqrt{1-4 \omega_0^2/\gamma^2}$.
In the limit $t \gg \tau$ Eqn.~\ref{e} reduces to $\ddot x = -
\gamma \dot x   -\omega_0^2 \sqrt{\tau/t} x $. In this limit we
have a stretched exponential solution
\begin{equation}
\label{g}
x(t)  \approx (x_0/2) \alpha_{+} \exp{( -\sqrt{t/t_0})}
\end{equation}
where $t_0 = \gamma^2/(\omega_0^4\tau ) = (27/2)\eta^2 D/(a_{MB}^2
E^2)$.  Significantly, $t_0$ does not depend on the initial
displacement $x$ or the initial applied force $F$.

We can compare these predictions to the experiment.  As mentioned
above, using $F_{\rm max}=0.75$~nN and $E \approx 15$~Pa,
$\xi_0 \approx 4a$.  The discussion in Sec.~\ref{experiment}
makes clear that neither $F_{\rm max}$ nor this inferred $E$
are the proper values for Eqn.~\ref{xi0}, but on the other
hand their ratio is what is needed to compute $\xi_0$ and it is
precisely this ratio that is directly measured in the experiments of
Sec.~\ref{experiment}.  We estimate $D$ as the short-time diffusion
coefficient, $D\approx k_B T/6 \pi \eta a = 0.064$~$\mu$m$^2$/s.
This approximation using the dilute-limit value is imperfect due
to hydrodynamic interactions which reduce $D$ at larger volume
fractions \cite{pusey83,snook83,beenakker83b,beenakker83}, but we
are mainly seeking the right order of magnitude.  Using this $D$ and
$\xi_0$ we find $\tau \approx 100$~s.  

The drag force acting on the magnetic bead is not due to the
viscosity $\eta$ of the solvent (used to calculate $D$) but rather
the effective viscosity of the medium, which is $\approx 50$ times
larger at these volume fractions \cite{cheng02}.  To calculate $t_0$
we use the more correct value of $E$ estimated from the data of
Fig.~\ref{fig-spring} as discussed in Sec.~\ref{experiment}.  Using
$a_{MB}=2.25$~$\mu$m and $E=0.72$~Pa we get $t_0 = 1$~ms.
This is too small by a factor of $\sim 200$ from the experimental
data (Fig.~\ref{fig-sqrt}).  Likewise, given $\tau \approx 100$~s,
we would expect to see the asymptotic (stretched exponential)
behavior for $\sqrt(t) \gg 10$ in Fig.~\ref{fig-sqrt}:  that we
see it at earlier time scales suggests that our estimate for $\tau$
is too large.

We thus reconsider the correct value of $D$.  In our model, we
assume $D$ is the diffusion coefficient for strain.  In practice,
individual colloidal particles do not need to move significant
distances for the strain to diffuse.  Much as a dislocation can
move rapidly through a crystalline lattice while individual
particles stay close to their lattice sites, a slight motion of a
particle ($\Delta r < a$) changes the strain over a neighborhood
$\sim a$ in scale.  If we assume that particles diffusing a
distance of $a/20$ is sufficient for the strain to diffuse a
distance $a$, then $D$ becomes 400 times larger.  This decreases
$\tau$ to 0.2~s and increases $t_0$ to 0.4~s, bringing our model
into more reasonable agreement with the data.  The distance
$a/20$ is smaller than the cage size (which is about $a/3$)
\cite{weeks02sub}.

\section{Conclusions}

We have used magnetic beads to locally perturb a dense colloidal
sample at volume fractions $\phi < \phi_g$, close to the colloidal
glass transition.  The magnetic beads have a linear relationship
between the applied force and their displacement, and the strain
field around the beads is well-described as that of a homogeneous
elastic medium subject to a point force.  The Poisson ratio is
$\sigma=1/2$, consistent with a sample that conserves its total
volume when a stress is applied.  Not surprisingly, the Young's
modulus describing the elastic medium grows as the glass transition
is approached.  The growth is consistent with power-law in $(\phi_c
- \phi)$, where $\phi_c = \phi_{rcp} > \phi_g$.

When the bead is moved away from its equilibrium position and the
force is removed, we observed the subsequent relaxation to the
equilibrium position.  This relaxation behaves as a stretched
exponential, $x \sim \exp(-(t/t_0)^{1/2})$.  This agrees with
a model that assumes the stress can diffuse away to infinity:
thus, while the particle is moving back to $x=0$, the effective
spring constant acting on the particle is also diminishing.
The experimental time scales suggest that this diffusion is rapid,
occurring faster than the particles themselves diffuse.  This is
likely due to the relatively small displacements of particles
needed to change the strain.

We thank R.~E.~Courtland, S.~A.~Koehler, K.~S.~Schweizer,
and M.~Wyart for helpful discussions.  We thank A.~Schofield
for providing our colloidal samples.  The work of D.~A., D.~S.,
P.~H., and J.~H. was supported by NASA (NAG3-2284).  The work of
E.~R.~W. was supported by NSF (CHE-0910707).


%
\end{document}